\documentclass{IEEEtran}
\usepackage{cite}
\usepackage{amsmath,amssymb,amsfonts}
\usepackage{graphicx}
\usepackage{textcomp,nicefrac}
\usepackage{siunitx}
\usepackage{xspace}
\usepackage{circledsteps}
\usepackage{tikz}
\usetikzlibrary{fit, patterns.meta}
\usepackage{pgfplots}
\pgfplotsset{compat=1.18} 
\usepackage{placeins}
\usepackage[caption=false,font=footnotesize]{subfig}
\usepackage{hyperref}
\usepackage{cleveref} 
\usepackage{orcidlink}

\usepackage[switch]{lineno}

\definecolor{CUDblack}{rgb}{0,0,0}
\definecolor{CUDorange}{rgb}{0.9,0.6,0}
\definecolor{CUDsky}{rgb}{0.35,0.70,0.90}
\definecolor{CUDgreen}{rgb}{0,0.6,0.50}
\definecolor{CUDyellow}{rgb}{0.95,0.90,0.25}
\definecolor{CUDblue}{rgb}{0,0.45,0.70}
\definecolor{CUDverm}{rgb}{0.8,0.4,0}
\definecolor{CUDred}{rgb}{0.8,0.6,0.7}

\newcommand{\belletwo}{Belle~II\xspace}
\newcommand{\gdl}{\texttt{GDL}\xspace}
\newcommand{\qkeras}{\texttt{QKERAS}\xspace}

\newcommand{\gnnetm}{\texttt{GNN-ETM}\xspace}
\newcommand{\icnetm}{\texttt{ICN-ETM}\xspace}
\newcommand{\vivado}{AMD~Vivado~2024.2\xspace}
\newcommand{\vitis}{AMD~Vitis~2024.2\xspace}
\newcommand{\modelsim}{ModelSim~2023.4\xspace}

\newcommand{\qnm}[2]{\texttt{Q#1.#2}\xspace}
\newcommand{\radiantarmadillo}{\emph{Armadillo}\xspace}
\newcommand{\dailysunset}{\emph{Sunset}\xspace}

\def\BibTeX{{\rm B\kern-.05em{\sc i\kern-.025em b}\kern-.08em
T\kern-.1667em\lower.7ex\hbox{E}\kern-.125emX}}
\begin{document}
\title{Commissioning and Low Latency Operation of the Graph Neural Network Electromagnetic Calorimeter Trigger at the Belle II Experiment}
\author{
  \IEEEauthorblockN{M.~Neu\textsuperscript{1}\,\orcidlink{0000-0002-4564-8009}},
  \IEEEauthorblockN{F.~Baptist\textsuperscript{1}\,\orcidlink{0009-0004-5007-5729}},
  \IEEEauthorblockN{I.~Haide\textsuperscript{1}\,\orcidlink{0000-0003-0962-6344}},
  \IEEEauthorblockN{Y.~Unno\textsuperscript{2}\,\orcidlink{0000-0003-3355-765X}},
  \IEEEauthorblockN{T.~Ferber\textsuperscript{1}\,\orcidlink{0000-0002-6849-0427}},
  \IEEEauthorblockN{J.~Becker\textsuperscript{1}\,\orcidlink{0000-0002-5082-5487}},
  \IEEEauthorblockN{K.~Arai\textsuperscript{3}\,\orcidlink{0009-0009-9301-8915}},
  \IEEEauthorblockN{Y.-T.~Lai\textsuperscript{4}\,\orcidlink{0000-0001-9553-3421}},
  \IEEEauthorblockN{T.~Koga\textsuperscript{4}\,\orcidlink{0000-0002-1644-2001}},
  \IEEEauthorblockN{M.~Maushart\textsuperscript{5}\,\orcidlink{0009-0004-1020-7299}},
  \IEEEauthorblockN{H.~Nakazawa\textsuperscript{6}\,\orcidlink{0000-0003-1684-6628}},
  \IEEEauthorblockN{V.~Savinov\textsuperscript{3}\,\orcidlink{0000-0002-9184-2830}},
  and \IEEEauthorblockN{K.~Unger\textsuperscript{1}\,\orcidlink{0000-0001-7378-6671}}
  \\
  \IEEEauthorblockA{\textsuperscript{1}Karlsruhe Institute of Technology (KIT), Karlsruhe, Germany}
  \IEEEauthorblockA{\textsuperscript{2}Hanyang University, Seoul, South Korea}
  \IEEEauthorblockA{\textsuperscript{3}University of Pittsburgh, Pittsburgh, United States}
  \IEEEauthorblockA{\textsuperscript{4}High Energy Accelerator Research Organiization (KEK), Tsukuba, Japan}
  \IEEEauthorblockA{\textsuperscript{5}Universit\'{e} de Strasbourg, Strasbourg, France}
  \IEEEauthorblockA{\textsuperscript{6}National Taiwan University, Taipei, Taiwan}
\thanks{The authors would like to thank the \belletwo collaboration for useful discussions and suggestions on
how to improve this work.
The training of the GNN-models was performed on the TOpAS GPU cluster at the Scientific
Computing Center (SCC) at the Karlsruhe Institute of Technology (KIT).}
}

\maketitle

\begin{abstract}
We present the commissioning and operation of the Graph Neural Network Electromagnetic Calorimeter Trigger Module (GNN-ETM) of the Belle II experiment at the SuperKEKB collider. The GNN-ETM processes calorimeter trigger cells as graph nodes to perform clustering and feature extraction. We fully integrate the system with the successive stages of the first-level trigger, develop slow-control drivers, and add online monitoring capabilities. We optimise the existing FPGA-based architecture through hardware–algorithm co-design, achieving an overall system latency of 1.053~us. Our hardware implementation is validated through register-transfer-level simulations, achieving bit-accurate agreement with the offline reference model. Online monitoring enables the measurement of instantaneous trigger rates, providing a quantitative basis for trigger-level performance studies. In summary, we report on the GNN-ETM as a fully operational, low-latency trigger module with online control and monitoring capabilities, compatible with the latency requirements of the \belletwo first-level trigger system.
\end{abstract}


\section{Introduction}
\label{sec:introduction}
\IEEEPARstart{G}{raph} Neural Networks (GNNs) have received increasing attention in the domain of high-energy physics for applications such as track reconstruction~\cite{reuter:2025}, particle tracking~\cite{elabd:2022}, hit cleanup~\cite{heine:2026:a}, calorimeter clustering~\cite{wemmer:2023}, jet tagging~\cite{qu:2020}, and event classification~\cite{shlomi:2021}.
So far, the majority of these studies have been performed in an offline environment, that is, without systematic constraints on throughput or latency.
GNNs for particle physics applications have nonetheless been deployed on Field-Programmable Gate Arrays (FPGAs) in prototypical studies for some time~\cite{heine:2026:a,que:2025,Dittmeier:2025nlh,neu:2025:a,que:2024,huang:2023}.
Recently, we presented the first GNN-based reconstruction algorithm implemented on FPGAs within the readout infrastructure of a collider experiment trigger system~\cite{haide:2026}.

The \belletwo experiment is located at the SuperKEKB~\cite{Akai:2018mbz} electron-positron collider in Tsukuba, Japan~\cite{Belle-II:2010dht}.
At SuperKEKB, \SI{4}{\giga\electronvolt} positrons collide with \SI{7}{\giga\electronvolt} electrons at a center-of-mass energy of approximately \SI{10.58}{\giga\electronvolt}.
At bunch crossing rates of up to \SI{254.4}{\mega\hertz}, defined by the SuperKEKB bunch filling pattern, reading out the full detector for each crossing is unfeasible due to bandwidth and storage limitations.
\belletwo therefore employs a trigger system that acts as an online filter, selecting potentially interesting events during operation based on a reduced subset of the detector signals.
The first-level~(L1) trigger operates under hard real-time requirements and is implemented as a pipeline of multiple FPGAs, each providing a custom compute module for a specific purpose.
One stage of this pipeline is the electromagnetic calorimeter (ECL) trigger, which clusters the energy depositions recorded in the calorimeter.

The \gnnetm, a real-time GNN-based calorimeter clustering algorithm implemented on FPGA, is one such module within the \belletwo ECL trigger~\cite{haide:2026}.
It currently operates in parallel to the existing ECL trigger system.
It is not used for trigger decision, but records debug data for further development.
The system supports a steady-state throughput of \num{8} million detector snapshots per second at an end-to-end latency of \SI{3.168}{\micro\second}.
While the module is real-time compliant, it is neither connected to the successive stages of the \belletwo L1 trigger system nor does its latency meet the hard real-time deadline of \SI{1.067}{\micro\second} required to contribute to the L1 trigger decision.

To overcome these limitations, this work extends the original \gnnetm of ref.~\cite{haide:2026} with the following contributions:
\begin{itemize}
    \item We introduce model compression and physical design optimisations, reducing the end-to-end latency from \SI{3.168}{\micro\second} to \SI{1.053}{\micro\second} to meet the \belletwo L1 trigger requirement, enabling the \gnnetm to actively contribute to the L1 trigger decision for the first time.
    \item We implement hardware modules that generate binary flags, so-called trigger bits, which are used in the global trigger decision, making the system compatible with the full L1 trigger chain.
    \item We implement real-time monitoring of trigger rates measured at the last stage of the L1 trigger system, demonstrating complete integration and enabling physics analyses on the developed trigger system by comparing \gnnetm with the existing \icnetm.
    \item We have commissioned and operated the improved \gnnetm in the \belletwo experiment since December 2025, validating its performance under nominal data-taking conditions.
\end{itemize}
\section{Background}
\label{sec:background}

\begin{figure*}[!ht]
    \centering
    \includegraphics[width=\linewidth]{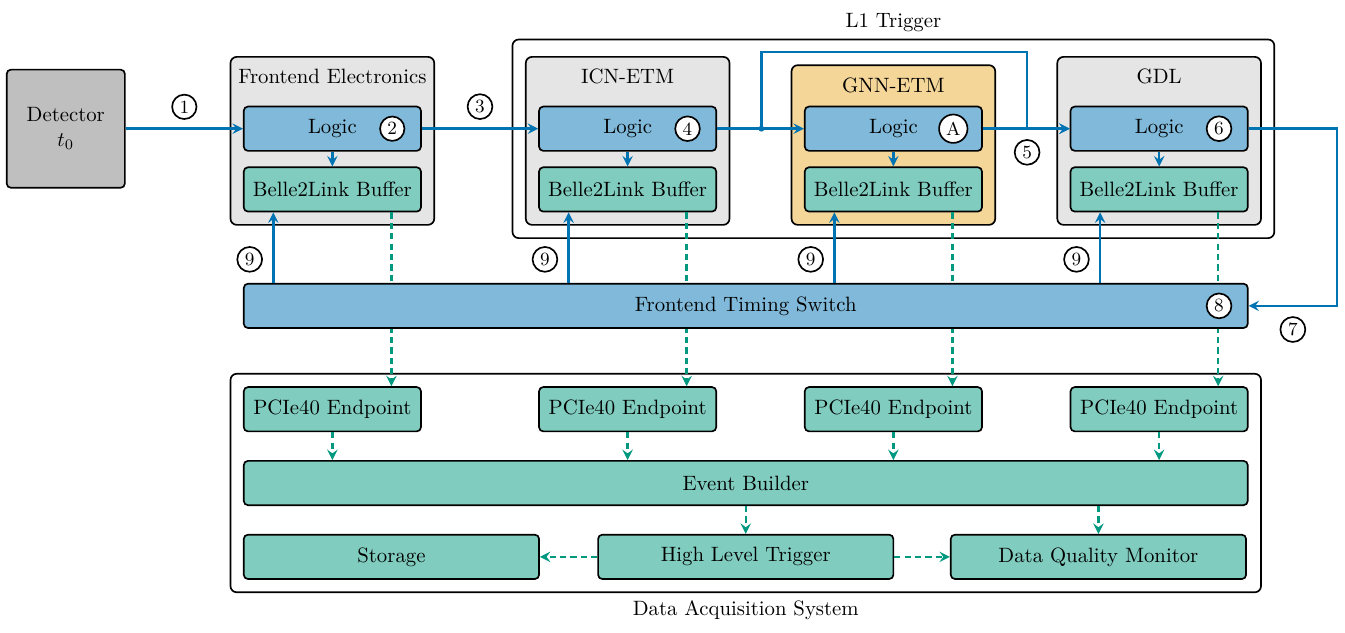}
    \caption{Simplified overview of the ECL L1 trigger system at the \belletwo Experiment.
    Components of the L1 trigger system that must satisfy hard real-time constraints are highlighted in blue, and components of the DAQ system that must satisfy soft real-time constraints are highlighted in green.
    The \gnnetm developed in this work is highlighted in orange.
    Numbered circles label selected components and connections referenced in the text.
    $t_0$ denotes the time of a detected bunch crossing.
    Adapted from ref.~\cite{haide:2026}. A detailed overview of the DAQ system is given in ref.~\cite{Yamada:2015xjy}.}
    \label{fig:triggersystem}
\end{figure*}

The data acquisition (DAQ) system at the \belletwo Experiment supports a maximum event readout rate of \SI{30}{\kilo\hertz}.
To reduce the computational load, a L1 trigger system is employed~\cite{Lai:2025gac}.
This trigger operates synchronously with the detector frontend readout at \SI{127.216}{\mega\hertz}, which is approximately half the bunch-crossing rate.
Detector snapshots are processed strictly sequentially.
To prevent buffer overflows in the DAQ readout system, the hard real-time latency budget, including data transfer, preprocessing, and synchronisation, is limited to \SI{5.0}{\micro\second}.

The \belletwo first-level trigger system comprises dedicated subtriggers for the participating subdetectors.
The task of the ECL trigger is to identify energy depositions for the global decision logic~\cite{kim:2017}.
\Cref{fig:triggersystem} depicts a simplified schematic of the ECL L1 trigger system.
For clarity, the connection to the Global Reconstruction Logic, where a matching between the ECL trigger and the trigger of other subdetectors is performed, is omitted.
The ECL detector is composed of 8736 thallium-doped caesium iodide crystals, which are read out via the Frontend Electronics~\Circled{1}~\cite{shwartz:2017}.
On the Frontend Electronics, crystals are first summed in the analog domain using $4\times4$ crystal groups.
A waveform fit is then performed to extract the signal amplitude and timing relative to the global revolution clock signal~\Circled{2}.
The resulting preprocessed groups of crystals, referred to as Trigger Cells (TCs), are uniquely identified by their positions within the detector and forwarded to the L1 trigger~\Circled{3}.
For the ECL L1 trigger, three modules are shown in the figure.

The first module is the \icnetm, an isolated clustering logic implemented in the Isolated Cluster Number ECL Trigger Module~\cite{cheon:2002}, which aggregates the information from all 576 TCs supplied by the Frontend Electronics~\Circled{4}.
Its purpose is to identify energy clusters and generate trigger bits.

The second module is the \gnnetm~\Circled{A}, receiving a copy of the ECL data from the \icnetm.
It runs in parallel to the \icnetm, also reconstructing energy clusters and generating trigger bits.

The \icnetm is connected to the Global Decision Logic~(\gdl) via optical fiber~\Circled{5}.
As part of this work, the connection between \gnnetm and \gdl is established and evaluated.
The \gdl aggregates trigger bits from all subdetectors~\Circled{6} to generate the global L1 trigger signal~\Circled{7}.
Trigger bits are Boolean flags classifying the current detector snapshot.
As an example, the two-cluster (C2) trigger bit on \icnetm is true if at least two clusters in the inner region of the ECL detector have been found with a per-cluster threshold of at least \SI{100}{\mega\electronvolt}.
The combination of trigger bits on \gdl is a combinatorial Boolean equation joining all trigger bits.
Inverted trigger bits may act as veto signals, for example, to suppress beam background.

The global L1 trigger signal is finally sent to the Frontend Timing Switch~\Circled{8}, which distributes the trigger signal to all modules in the Frontend Electronics as well as all modules in the trigger system~\Circled{9}.
The Belle2Link Buffers store full-resolution and trigger data for a specified time period.
When a trigger signal is received, data is sent to the PCIe40 Endpoints of the DAQ system, where the individual packets are assembled into a common format in the Event Builder.
The events are then processed by the second filtering stage in the \belletwo trigger system, the High Level Trigger (HLT), based on a CPU farm with soft real-time constraints.
Finally, events that pass the HLT are persistently stored on disk for later analysis.
All other data are lost and cannot be recovered.

Operating the \gnnetm in the position described in \Cref{fig:triggersystem} imposes the following requirements on the system, based on the analysis in ref.~\cite{haide:2026}:
\begin{enumerate}
    \item The system must exhibit deterministic latency to satisfy hard real-time deadlines.
    \item The critical-path latency \Circled{1} $\rightarrow$ \Circled{9} must not exceed $R_\mathrm{L} = \SI{5.0}{\micro\second}$, a constraint imposed by the finite depth of the data buffers on the vertex detector frontend electronics modules.
    In practice, only a small fraction of this latency is available for a subsystem trigger module.
    For the \belletwo ECL L1 trigger system, the latency \Circled{4} of the module replacing the current \icnetm must not exceed \SI{1.067}{\micro\second}\footnote{The value here differs from the previously reported value in ref.~\cite{haide:2026}. The reason is that we do not swap the order of \icnetm and \gnnetm in the ECL L1 trigger system in this work, as it incurs additional implementation overhead.}.
    \item The system must sustain the full input rate of the respective subdetector, which for the \belletwo ECL L1 trigger amounts to $R_\mathrm{th} = \SI{8}{\mega\hertz}$.
    \item The system must maintain \SI{100}{\percent} uptime, as any disruption halts the entire experiment for the duration of the fault.
\end{enumerate} 
\section{GNN-ETM Architecture}
\label{sec:architecture}
\begin{figure}[h]
    \centering
    \includegraphics[width=\linewidth]{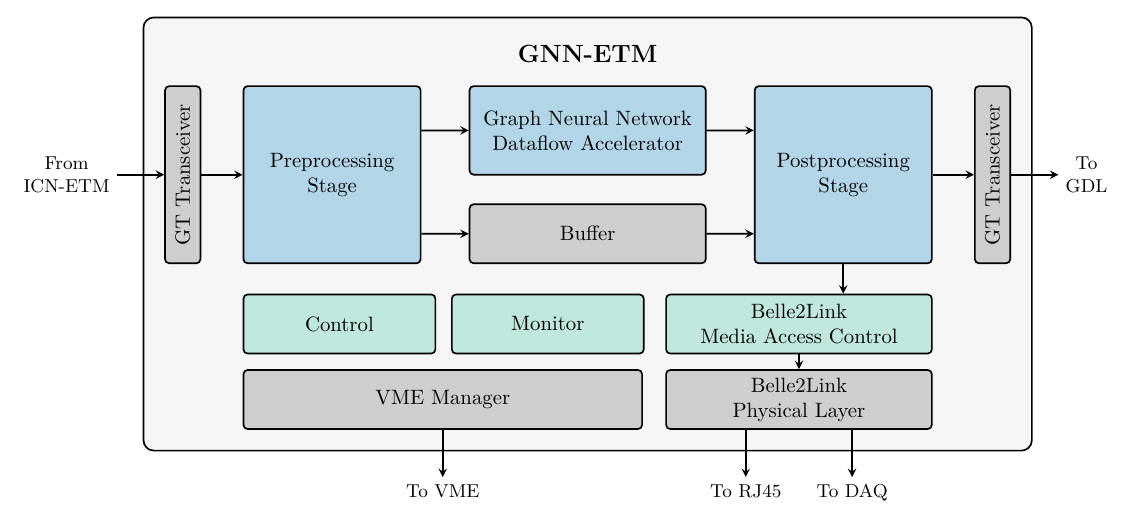}
    \caption{Overview of the \gnnetm system architecture. Existing components of the
    base firmware is shown in grey. Modules introduced in this work are coloured:
    modules on the critical path of the trigger system are blue, the remaining modules
    are green. Adapted from ref.~\cite{haide:2026}.}
    \label{fig:architecture}
\end{figure}

The \gnnetm is realised as an FPGA-based architecture on the fourth generation of the
Universal Trigger Board (UT4). It interfaces with the up- and downstream modules in the
\belletwo L1 trigger chain via AXI-Stream~\cite{ARM:IHI0051B}.
These interfaces are carried over optical links realised with gigabit transceivers.
For slow control and monitoring, the FPGA communicates with a general-purpose CPU via the
Versa Module Eurocard (VME) bus~\cite{IEEE:1014-1987}. Debug data is transmitted via the Belle2Link
physical layer protocol~\cite{sun:2012,Yamada:2015xjy}.

A system overview of the \gnnetm is shown in \Cref{fig:architecture}. It comprises three
submodules on the critical path: the preprocessing stage, the GNN dataflow accelerator,
and the postprocessing stage. Further submodules provide monitoring, control, and the
Belle2Link media access control, which handles the interfaces to VME and to the Belle2Link
for debugging.
Compared to the \gnnetm described in ref.~\cite{haide:2026}, we introduce the following
modifications to the architecture.

First, we adapt the Chisel-based~\cite{bachrach:2012} preprocessing
and postprocessing stages to include clock domain crossings between the submodules,
enabling a user-defined system frequency $f_\mathrm{sys}$ for the GNN dataflow accelerator.
We choose synchronous clock domain crossings to mitigate boundary effects and to minimise the latency overhead of the crossing.

Second, we fully integrate the \gnnetm with the slow control system of the \belletwo
experiment~\cite{konno:2015,kim:2021}. This integration enables automated configuration, error handling, and
monitoring. For example, at the start of every run the \gnnetm parameters are read from a
database via Network Shared Memory~2 (NSM2)~\cite{nakao:1999}
and written directly to the firmware register map through VME. The same interface is used
for real-time monitoring of the \gnnetm trigger rates.

\begin{figure}[h]
    \centering
    \includegraphics[width=\linewidth]{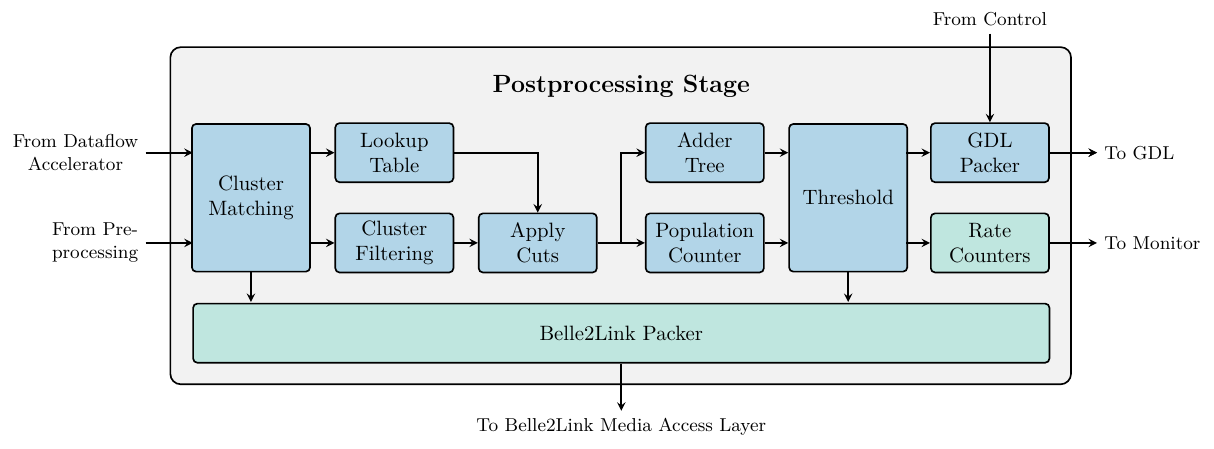}
    \caption{Overview of the \gnnetm postprocessing stage. Latency-critical submodules are blue; the remaining submodules are green. The external interfaces of this submodule
    are also shown in \Cref{fig:architecture}.}
    \label{fig:postprocessing}
\end{figure}

Third, we extend the postprocessing stage with the generation of trigger bits and a
monitoring counter for each trigger bit. An overview is given in \Cref{fig:postprocessing}.

The postprocessing stage extracts the final trigger bits from the \gnnetm cluster predictions.
It comprises ten submodules, eight of which reside in the critical timing path.

First, the cluster matching module synchronises the cluster features from the GNN dataflow accelerator with the corresponding TCs from the preprocessing stage.
Since TCs are processed strictly in order, the matching problem reduces to concatenating synchronised trigger-cell and cluster features.
Second, a lookup table retrieves the static TC information required by subsequent modules, such as the position of each TC.
In parallel with the lookup, the cluster filtering submodule applies the condensation point selection mask, setting all inactive clusters to zero so that only valid clusters propagate through the pipeline.
Third, energy and multiplicity cuts are applied to form the individual trigger bits:
parallel adder trees compute energy-sum trigger bits, which calculate the total energy deposited in the ECL, while
population counters capture the cluster-counting trigger bits, which count the number of clusters above a certain energy threshold in the ECL.
Finally, all trigger bits are packed for transmission to the \gdl module.

Separately, two monitoring modules process the data further.
The raw rate of each trigger bit is tracked using \SI{32}{\bit} counters, which are read out via the slow control interface over VME.
In addition, a copy of all trigger bits is sent to the Belle2Link media access control system for further debugging of the \gnnetm in operation.
\section{Deployment}
\label{sec:deployment}

For deploying the GNN algorithm on the \gnnetm architecture, we use the approach previously described in ref.~\cite{haide:2026}.
The deployed GNN model is the CaloClusterNet, a dynamic GNN model, based on the GravNet~\cite{qasim:2019} layer and the Object Condensation algorithm~\cite{Kieseler:2020wcq}.
Because the baseline version of the CaloClusterNet, hereafter referred to as \radiantarmadillo CaloClusterNet, does not meet the latency requirements of the \belletwo L1 trigger system, we apply three optimisation steps to reduce the overall latency of the system while minimising the loss of algorithmic performance of the model.
As the target FPGA on the UT4, the AMD~Ultrascale~XCVU190 is chosen.
For the deployment and evaluation, we use \vitis~\cite{vitis:2024:2} and \vivado~\cite{vivado:2024:2}.
In the following, we describe four design iterations:
Design iteration~\Circled{1} describes the baseline implementation from ref.~\cite{haide:2026} using the \radiantarmadillo CaloClusterNet.
Design iteration~\Circled{2} describes the deployment of the improved version after model compression in \Cref{sec:modelcompression}.
Design iteration~\Circled{3} describes the deployment after manual floorplanning in \Cref{sec:floorplanning}.
Design iteration~\Circled{4} describes the deployment after DSP-level optimisations in \Cref{sec:dsp}.

\subsection{Model Compression}
\label{sec:modelcompression}

In the first design iteration, we aim to reduce the latency of the baseline \radiantarmadillo CaloClusterNet through optimised quantisation-aware training.
We use \qkeras~\cite{coelho:2021} with the adaptation from refs.~\cite{gnnetm-software,qgravnet,qkeras-code} to implement and train the network.
Our \dailysunset CaloClusterNet incorporates the following optimisations in comparison to the original \radiantarmadillo CaloClusterNet.

First, we reduce the number of GravNet blocks from two to one, significantly decreasing the network's complexity.

Second, we perform a manual hyperparameter search for layerwise heterogeneous fixed-point quantisation.
To reduce the hardware resource utilisation on the FPGA, we impose a hard limit of \SI{8}{\bit} per network layer.
Quantisation is uniform within each layer.
Power-of-two quantisation is chosen to enable efficient requantisation between adjacent neural network layers.

Third, we apply stochastic rounding and add quantisation noise during training to improve the trainability of the neural network under stricter quantisation schemes by reducing the bias imposed by the quantisation scheme~\cite{gupta:2015,liu:2019}.

\begin{figure}[h]
    \centering
    \includegraphics[width=\linewidth]{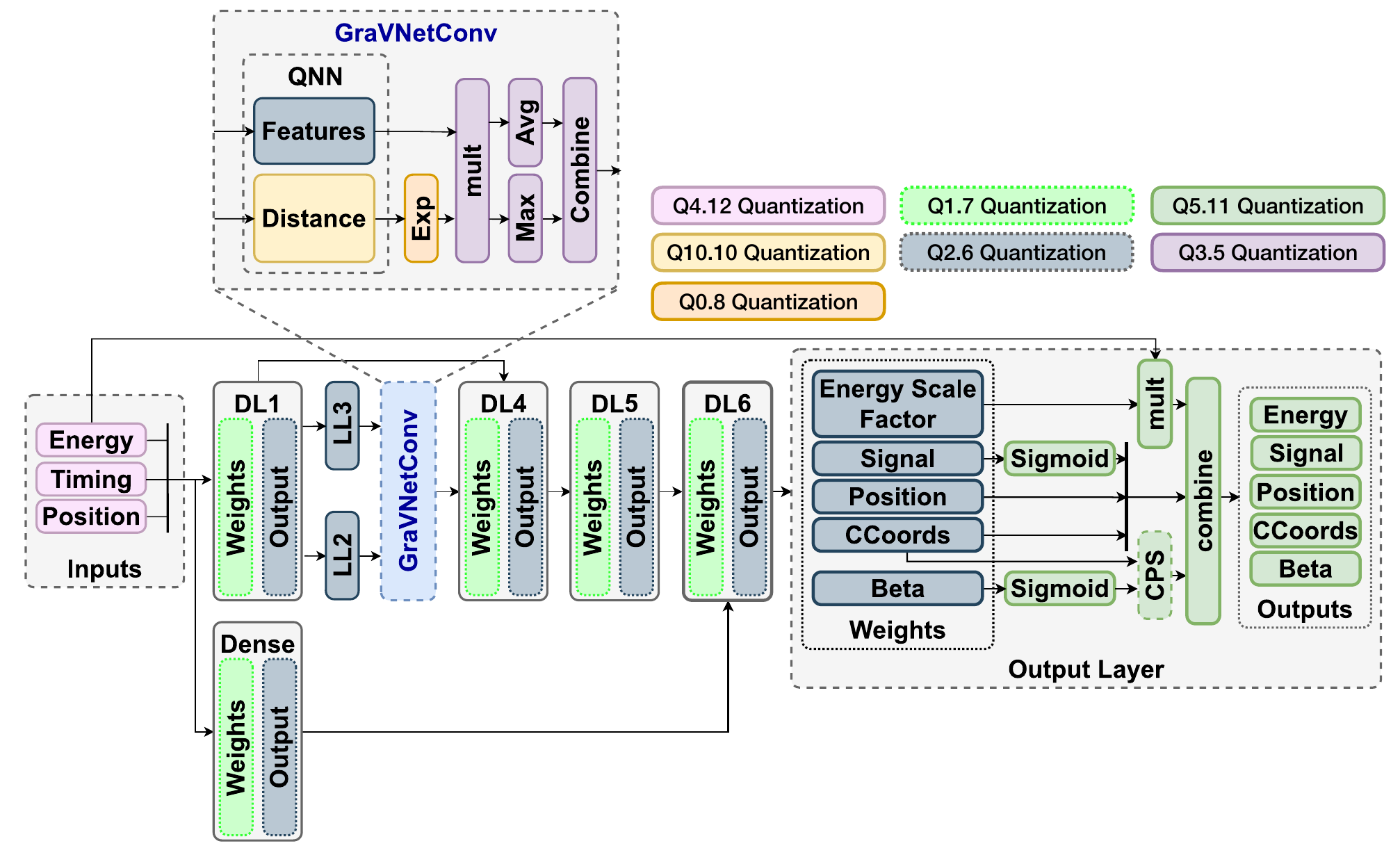}
    \caption{
    Our compressed \dailysunset CaloClusterNet neural network architecture. This model architecture is derived from ref.~\cite{haide:2026}.}
    \label{fig:model}
\end{figure}

The model topology of our resulting network is shown in \Cref{fig:model}.
Most values are quantized to \qnm{1}{7} and \qnm{2}{6}, meaning that most values have to lie in the range $[-1, 1]$ and $[-2, 2]$ respectively.
At the interfaces, we keep the \qnm{4}{12} and \qnm{5}{11} quantisation to retain the full resolution.
In comparison to the \radiantarmadillo CaloClusterNet, the \dailysunset CaloClusterNet does not yield a significantly lower algorithmic performance.

\subsection{Floorplanning}
\label{sec:floorplanning}

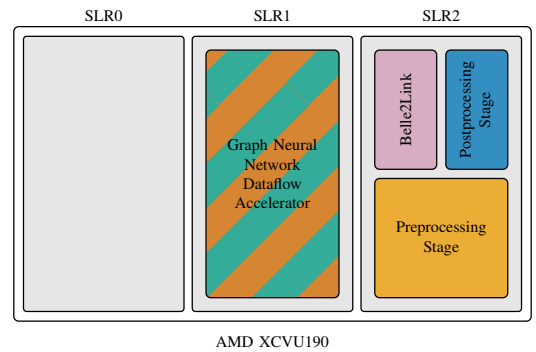
\begin{figure}[h]
    \centering
    \resizebox{.8\linewidth}{!}{
\providecommand{\pblock}[7]{%
  \draw[fill=#3, draw=#4, thick, rounded corners=3pt]
    (#1,#2) rectangle ({#1+#5},{#2+#6});
  \node[align=center] at ({#1+0.5*#5},{#2+0.5*#6}) {#7};%
}
\providecommand{\pblockpattern}[8]{%
  \path[rounded corners=3pt,
        fill=#3,
        postaction={pattern={Lines[angle=45, distance=24pt, line width=12pt]},
                    pattern color=#4},
        postaction={draw=#5, thick}]
    (#1,#2) rectangle ({#1+#6},{#2+#7});
  \node[align=center] at ({#1+0.5*#6},{#2+0.5*#7}) {#8};%
}

\begin{tikzpicture}
  \draw[thick, fill=gray!20, rounded corners=2pt] (0.00,0) rectangle (3.50,6);
  \draw[thick, fill=gray!20, rounded corners=2pt] (3.68,0) rectangle (7.18,6);
  \draw[thick, fill=gray!20, rounded corners=2pt] (7.36,0) rectangle (10.86,6);
  \node[anchor=south] at (1.75, 6.2) {SLR0};
  \node[anchor=south] at (5.43, 6.2) {SLR1};
  \node[anchor=south] at (9.11, 6.2) {SLR2};
  \node[draw, thick, rounded corners=3pt, fit={(0,0)(10.86,6)}, inner sep=6pt,
        label={[yshift=-5pt]below:{AMD XCVU190}}] {};
  \pblock{7.66}{0.3}{CUDorange!80}{CUDblack}{2.9}{2.6}  {Preprocessing \\ Stage}
  \pblock{7.66}{3.1}{CUDred!80}{CUDblack}{1.35}{2.6}    {\rotatebox{90}{Belle2Link}}
  \pblock{9.21}{3.1}{CUDblue!80}{CUDblack}{1.35}{2.6}   {\rotatebox{90}{\shortstack{Postprocessing\\Stage}}}
  \pblockpattern{3.98}{0.3}{CUDverm!80}{CUDgreen!80}{CUDblack}{2.9}{5.4}
    {Graph Neural \\ Network \\ Dataflow \\ Accelerator}
\end{tikzpicture}}
    \caption{Floorplan constraints of the \gnnetm for implementation with \vivado. Hierarchical modules are the same as in \Cref{fig:architecture}. Hierarchical modules that do not appear in the floor plan are not subject to any location constraints.}
    \label{fig:floorplan}
\end{figure}

In the second design iteration, we consider manual floorplanning to improve the design's routability and increase the maximum achievable design frequency $f_\mathrm{sys}$.
Because the AMD~Ultrascale~XCVU190 is composed of three identical Super Logic Regions (SLRs), manually floorplanning the architecture can drastically improve $f_\mathrm{sys}$, as netlist wires crossing these SLRs without further optimisation result in either congested routes or routing delays that dominate the path.
In ref.~\cite{haide:2026}, the GNN dataflow accelerator operates at \SI{127.216}{\mega\hertz}, while the preprocessing stage already runs at the doubled frequency of \SI{254.432}{\mega\hertz}.
Through manual floorplanning, we now also operate the GNN dataflow accelerator at \SI{254.232}{\mega\hertz}, applying the floorplanning constraints described in \Cref{fig:floorplan}.
Notably, using the previous model compression in \Cref{sec:modelcompression}, we can implement the GNN dataflow accelerator on a single SLR.

\subsection{DSP Mapping}
\label{sec:dsp}

Due to the multiplication-heavy design of our dataflow accelerator, we usually maximise the utilisation of the hard DSP blocks on the FPGA, as they tend to achieve a better performance than realising multiplication in distributed logic.
However, during implementation, we observed that the distribution of the hard-macro DSPs on the FPGA results in suboptimal placement and, in turn, routing congestion, making timing closure more difficult.
In addition, hard-macro DSPs on the AMD~Ultrascale fabric require up to three internal registers for full pipelining, resulting in some latency overhead.
Thus, in a final optimisation step, we disable all hard-macro DSPs via the corresponding configuration options in \vitis and \vivado.
\section{Performance Analysis}
\label{sec:performance}
We measure the end-to-end system performance of \gnnetm in two different ways before commissioning the system in the \belletwo L1 trigger system.
First, we perform a cycle-accurate register-transfer-level simulation of the complete design in \modelsim~\cite{modelsim:2023:4}.
From this simulation, we validate functional correctness, verify that the throughput requirement is met, and derive the system's end-to-end latency in \Cref{sec:latency}.
Second, we implement the design on the UT4 board and validate that all timing constraints are met after place-and-route.
We analyse the \vivado report after implementation in \Cref{sec:systemresources}.

\subsection{Latency}
\label{sec:latency}
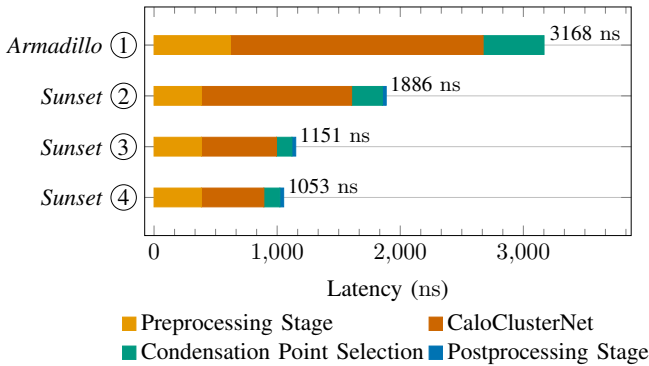
\begin{figure}[h]
    \centering
    \resizebox{\linewidth}{!}{ \begin{tikzpicture}
        \begin{axis}[
            xbar stacked,
            xlabel={Latency (\si{ns})},
            xmin=0, xmax=3800,
            ytick={1,2,3,4},
            yticklabels={\emph{Sunset} \Circled{4}, \emph{Sunset} \Circled{3}, \emph{Sunset} \Circled{2}, \emph{Armadillo} \Circled{1}},
            enlarge x limits=0.02,
            enlarge y limits=0.25,
            width=\linewidth,
            height=5.0cm,
            bar width=8pt,
            ymajorgrids = true,
            minor x tick num=5,
            legend columns=2,
            legend style={
                at={(0.5,-0.30)},
                anchor=north,
                draw=none,
                legend cell align=left,
            },
            legend image code/.code={%
                \draw[#1] (0cm,-0.1cm) rectangle (0.2cm,0.1cm);
            }
        ]
            \addplot+[style={CUDorange,fill=CUDorange,mark=none}] coordinates {(621,4) (385,3) (385,2) (385,1)};
            \addplot+[style={CUDverm,fill=CUDverm,mark=none}] coordinates {(2052,4) (1220,3) (610,2) (507,1)};
            \addplot+[style={CUDgreen,fill=CUDgreen,mark=none}] coordinates {(495,4) (250,3) (125,2) (130,1)};
            \addplot+[style={CUDblue,
                      fill=CUDblue,
                      mark=none},
                      nodes near coords,
                      point meta=explicit symbolic,
                      every node near coord/.append style={
                        anchor=west,
                        xshift=-1pt,
                        yshift=5pt,
                        text=black
                      }
            ] coordinates {(0,4)
                           (31,3) [\small{1886~\si{ns}}]
                           (31,2) [\small{1151~\si{ns}}]
                           (31,1) [\small{1053~\si{ns}}]};
            \node[anchor=west, xshift=-1pt, yshift=5pt, text=black] at (axis cs:3168,4) {\small 3168~\si{ns}};
            \addlegendentry{Preprocessing Stage}
            \addlegendentry{CaloClusterNet}
            \addlegendentry{Condensation Point Selection}
            \addlegendentry{Postprocessing Stage}
        \end{axis}
    \end{tikzpicture}}
    \caption{End-to-end latency for the complete inference chain on the UT4 with an AMD~Ultrascale~XCVU190 FPGA. Design iterations~\Circled{1}---\Circled{4} are presented.}
    \label{fig:latency}
\end{figure}

\Cref{fig:latency} depicts the latency of all four design iterations.
The baseline design in~\Circled{1} requires \SI{3168}{\nano\second} for the end-to-end inference, including the preprocessing stage.
The baseline design does not include generating trigger bits.
After applying the model compression in~\Circled{2}, the latency is reduced by \SI{1282}{\nano\second}.
Further optimisation of the floorplanning in~\Circled{3} reduces the latency by an additional \SI{735}{\nano\second}.
A final \SI{98}{\nano\second} are saved in~\Circled{4}, resulting in an end-to-end latency of the \gnnetm of \SI{1053}{\nano\second}, which is a \num{3.01}$\times$ reduction over the baseline version.
Breaking down the latency, \SI{385}{\nano\second} are used by the preprocessing stage, \SI{507}{\nano\second} by the CaloClusterNet, \SI{130}{\nano\second} by the Condensation Point Selection, and \SI{31}{\nano\second} by the postprocessing stage.

\subsection{System Resource Utilisation}
\label{sec:systemresources}
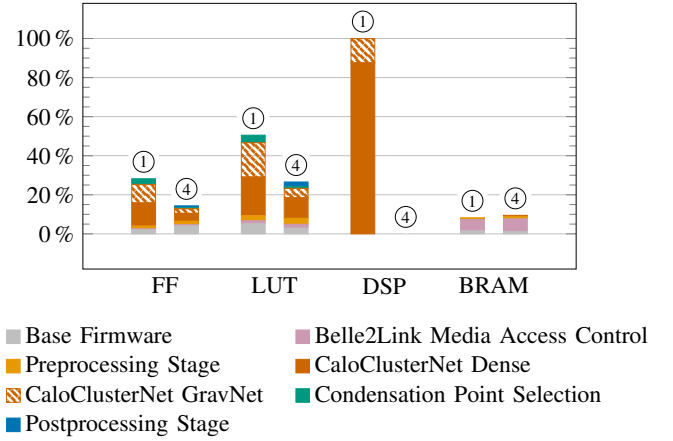
\begin{figure}[h]
    \centering
    \resizebox{\linewidth}{!}{  \begin{tikzpicture}
    \pgfplotsset{
      util axis/.style={
        width=\linewidth, height=5.5cm,
        ybar stacked, bar width=10pt, ymajorgrids=true,
        symbolic x coords={FF,LUT,DSP,BRAM},
        xtick=data, scaled y ticks=false,
        enlarge x limits=0.25, enlarge y limits=0.18,
        ymin=0, ymax=100,
        ytick={0,20,40,60,80,100},
        yticklabels={0\,\%,20\,\%,40\,\%,60\,\%,80\,\%,100\,\%},
        minor y tick num=3,
        major x tick style=transparent,
      }
    }
    \begin{axis}[
        util axis, bar shift=-9pt,
        legend columns=2,
        legend style={at={(0.5,-0.18)},anchor=north,draw=none,legend cell align=left,
            /tikz/every even column/.append style={column sep=0.4cm},
            /tikz/column 1/.style={anchor=base west, text width=4.6cm}},
        legend image code/.code={\draw[#1] (0cm,-0.1cm) rectangle (0.2cm,0.1cm);},
    ]
        \addplot+[style={CUDblack!25,fill=CUDblack!25,mark=none}] coordinates {(BRAM,1.56)(DSP,0.00)(FF,2.15)(LUT,5.25)};
        \addplot+[style={CUDred,fill=CUDred,mark=none}] coordinates {(BRAM,5.73)(DSP,0.00)(FF,0.40)(LUT,1.45)};
        \addplot+[style={CUDorange,fill=CUDorange,mark=none}] coordinates {(BRAM,0.95)(DSP,0.00)(FF,1.50)(LUT,2.69)};
        \addplot+[style={CUDverm,fill=CUDverm,mark=none}] coordinates {(BRAM,0.00)(DSP,87.56)(FF,11.79)(LUT,19.71)};
        \addplot+[style={CUDverm,mark=none},pattern={Lines[angle=-45,distance=2pt,line width=1pt]},pattern color=CUDverm] coordinates {(BRAM,0.00)(DSP,12.44)(FF,9.47)(LUT,17.59)};
        \addplot+[style={CUDgreen,fill=CUDgreen,mark=none}] coordinates {(BRAM,0.00)(DSP,0.00)(FF,2.91)(LUT,3.83)};
        \addplot+[style={CUDblue,fill=CUDblue,mark=none}] coordinates {(BRAM,0.00)(DSP,0.00)(FF,0.00)(LUT,0.00)};
        \addlegendentry{Base Firmware}
        \addlegendentry{Belle2Link Media Access Control}
        \addlegendentry{Preprocessing Stage}
        \addlegendentry{CaloClusterNet Dense}
        \addlegendentry{CaloClusterNet GravNet}
        \addlegendentry{Condensation Point Selection}
        \addlegendentry{Postprocessing Stage}
        \node[font=\scriptsize,anchor=south,xshift=-9pt,circle,fill=white,inner sep=0.5pt] at (axis cs:FF,28.22)  {\Circled{1}};        \node[font=\scriptsize,anchor=south,xshift=-9pt,circle,fill=white,inner sep=0.5pt] at (axis cs:LUT,50.52) {\Circled{1}};
        \node[font=\scriptsize,anchor=south,xshift=-9pt,circle,fill=white,inner sep=0.5pt] at (axis cs:DSP,100.0) {\Circled{1}};
        \node[font=\scriptsize,anchor=south,xshift=-9pt,circle,fill=white,inner sep=0.5pt] at (axis cs:BRAM,8.24) {\Circled{1}};
    \end{axis}

    \begin{axis}[
        util axis, bar shift=9pt,
        axis x line=none, axis y line=none, ymajorgrids=false,
        tick style={draw=none}, xtick=\empty, ytick=\empty,
    ]
        \addplot+[style={CUDblack!25,fill=CUDblack!25,mark=none}] coordinates {(BRAM,1.20)(DSP,0.00)(FF,4.10)(LUT,3.00)};
        \addplot+[style={CUDred,fill=CUDred,mark=none}] coordinates {(BRAM,6.50)(DSP,0.00)(FF,0.60)(LUT,1.90)};
        \addplot+[style={CUDorange,fill=CUDorange,mark=none}] coordinates {(BRAM,1.10)(DSP,0.00)(FF,1.80)(LUT,3.10)};
        \addplot+[style={CUDverm,fill=CUDverm,mark=none}] coordinates {(BRAM,0.70)(DSP,0.0)(FF,4.00)(LUT,10.50)};
        \addplot+[style={CUDverm,mark=none},pattern={Lines[angle=-45,distance=2pt,line width=1pt]},pattern color=CUDverm] coordinates {(BRAM,0.00)(DSP,0.0)(FF,2.40)(LUT,4.80)};
        \addplot+[style={CUDgreen,fill=CUDgreen,mark=none}] coordinates {(BRAM,0.00)(DSP,0.00)(FF,0.45)(LUT,0.85)};
        \addplot+[style={CUDblue,fill=CUDblue,mark=none}] coordinates {(BRAM,0.00)(DSP,0.00)(FF,1.00)(LUT,2.40)};
        \node[font=\scriptsize,anchor=south,xshift=9pt,circle,fill=white,inner sep=0.5pt] at (axis cs:FF,14.35)  {\Circled{4}};
        \node[font=\scriptsize,anchor=south,xshift=9pt,circle,fill=white,inner sep=0.5pt] at (axis cs:LUT,26.55) {\Circled{4}};
        \node[font=\scriptsize,anchor=south,xshift=9pt,circle,fill=white,inner sep=0.5pt] at (axis cs:DSP,0)     {\Circled{4}};
        \node[font=\scriptsize,anchor=south,xshift=9pt,circle,fill=white,inner sep=0.5pt] at (axis cs:BRAM,9.5)  {\Circled{4}};
    \end{axis}
  \end{tikzpicture}}
    \caption{Utilisation of system resources on the AMD~Ultrascale~XCVU190 FPGA for the \gnnetm with \radiantarmadillo~\Circled{1} and \dailysunset~\Circled{4} CaloClusterNet model.}
    \label{fig:utilisation}
\end{figure}

\Cref{fig:utilisation} depicts the system resource utilisation after place and route on the UT4 for the baseline design~\Circled{1} and the final design~\Circled{4}.
In comparison between the two versions, both flip-flops (FFs) and lookup tables (LUTs) are reduced by approximately \SI{50}{\percent}.
The main reason for this difference lies in the removal of one GravNet layer and the reduced precision of all Dense layers.
This effect also influences the utilisation of the successive Condensation Point Selection submodule.
Similarly, DSPs are now unused, and multiplications are mapped to distributed logic after applying the optimisation from \Cref{sec:dsp}.
A slight increase in Block RAM (BRAM) utilisation is observed due to the addition of trigger bits in the Belle2Link readout.
\section{Commissioning \& Operation}
\label{sec:operation}

To validate the \gnnetm, we commission the system in the \belletwo L1 trigger system as depicted in \Cref{fig:triggersystem}.
Compared with the commissioning of the previous system in ref.~\cite{haide:2026}, we develop slow-control and monitoring software compatible with the general \belletwo run control and add the upstream link to the \gdl.
Configurations, trigger rates, and monitoring flags are broadcast via NSM2, and additionally registered as process variables (PVs) in the \belletwo EPICS archiver database for later analysis~\cite{shankar:2015}.
Polling-sampling mode is used for logging \gnnetm PVs at a rate of \SI{1}{\hertz}.

\begin{figure}[h]
    \centering
    \includegraphics[width=\linewidth]{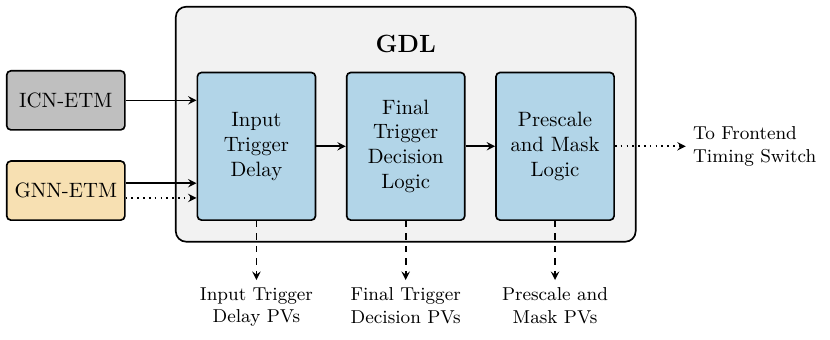}
    \caption{Interfaces between \gnnetm, \icnetm, and \gdl. Interfaces via gigabit transceivers are shown as solid arrows. Interfaces via twisted-pair cables are depicted as dotted arrows. Slow control interfaces are depicted as dashed arrows.}
    \label{fig:commissioning}
\end{figure}

\Cref{fig:commissioning} depicts the interfaces of \icnetm, \gnnetm, and \gdl in the \belletwo L1 trigger system.
Both \icnetm and \gnnetm are connected via gigabit transceivers (GTs) to the \gdl.
This connection via optical fibres enables the transmission of large packets of up to \SI{768}{\bit} per \SI{127.216}{\mega\hertz} system clock cycle, with a latency of approximately \SI{200}{\nano\second}.
To avoid this latency overhead, we add a single twisted-pair (LEMO) cable between \gnnetm and \icnetm.
This transmission interfaces directly with the high-speed-capable pins on the FPGA, avoiding error correction, synchronisation, and other physical-layer overhead present in the GT connections.

On the \gdl, trigger bits are received from all L1 trigger subsystems.
For simplicity, only \icnetm and \gnnetm are shown in the figure.
In general, trigger bits pass through three submodules:

First, the input trigger delay submodule checks the connection to the \gdl for liveness and measures the subtrigger latency via the \texttt{active} signal.
Variable-length shift registers are used to synchronise all incoming subtriggers based on the measured delay.
The \gdl implements rate counters for the \texttt{active} signal and all trigger bits after this stage, measuring the raw input trigger rate and storing it in the EPICS archiver database.
Trigger bits included in the \gdl are monitored via the Input Trigger Delay PVs.

Second, an optional veto (bitwise \texttt{AND} with the inverted veto signal) is applied to the input trigger signals and again recorded as Final Trigger Decision PVs.
Two vetoes used for the \gnnetm trigger signals are the injection veto and the Bhabha veto.
The injection veto suppresses beam-induced background.
The Bhabha veto suppresses the high-rate Bhabha scattering process.
Without applying these vetoes, both would dominate the resulting trigger rates.

Third, a prescale and mask value is applied to all trigger bits.
This effectively enables the run operators to disable single bits (masking) or to reduce the trigger rate of these bits by triggering only on the $N$th occurrence (where $N$ is the prescale factor).
The output of this stage is recorded as Prescale and Mask PVs.

In this work, \gnnetm trigger bits are monitored in the \gdl, but do not actively contribute to the trigger decision in the \belletwo L1 trigger system.
In the following, \gnnetm will be evaluated in runs.
A run is defined as a data-taking period during which the experiment and the accelerator configuration remain constant.
We differentiate between two run types.
Cosmic runs are data-taking periods without beam, whereas beam runs are data-taking periods in which the beams collide in the interaction point of the \belletwo Experiment.

\subsection{Latency Measurement}

We derive the end-to-end latency and the latency requirement for \gnnetm in a beam run. 
The system latency is measured by observing the histogram of the rising-edge clock counter implemented on the \gdl, recorded over the full run.
The observed input delay is denoted by $\hat{t}$.
The rising-edge clock counter continuously monitors each trigger bit and stores the delay value with a resolution of \SI{32}{\nano\second} for both the gigabit transceiver and the twisted-pair cable.
For the analysis, a histogram of arrival times is recorded over a full run using a dedicated clock counter module on the \gdl.
The upper acceptable input delay at the \gdl has been defined as 20 system clock cycles, or \SI{640}{\nano\second}, based on experimental evaluation during data acquisition (DAQ) stress tests.
The input delay is measured against an arbitrarily chosen reference time on the \gdl, which depends on the clock distribution architecture at \belletwo.

To relate a connection between $\hat{t}$ and the actual \gnnetm latency $\hat{t}_\mathrm{gnn}$, we apply the following offsets:
First, we apply the programmable delay offset $t_\mathrm{FAM}$ from the Frontend Analog Module to remove the effects of different run configurations.
Second, we apply an offset based on the difference between the simulated cycle-accurate latency $t_\mathrm{sim}$ and the \SI{95}{\percent} quantile $\hat{t}_{0.95}$ measured via the twisted-pair cable:

\begin{equation}
  \hat{t}_\mathrm{gnn} = \hat{t} + t_\mathrm{FAM} + t_\mathrm{sim} - \hat{t}_{0.95}
\end{equation}

\begin{figure}[h]
    \centering
    \includegraphics[width=\linewidth]{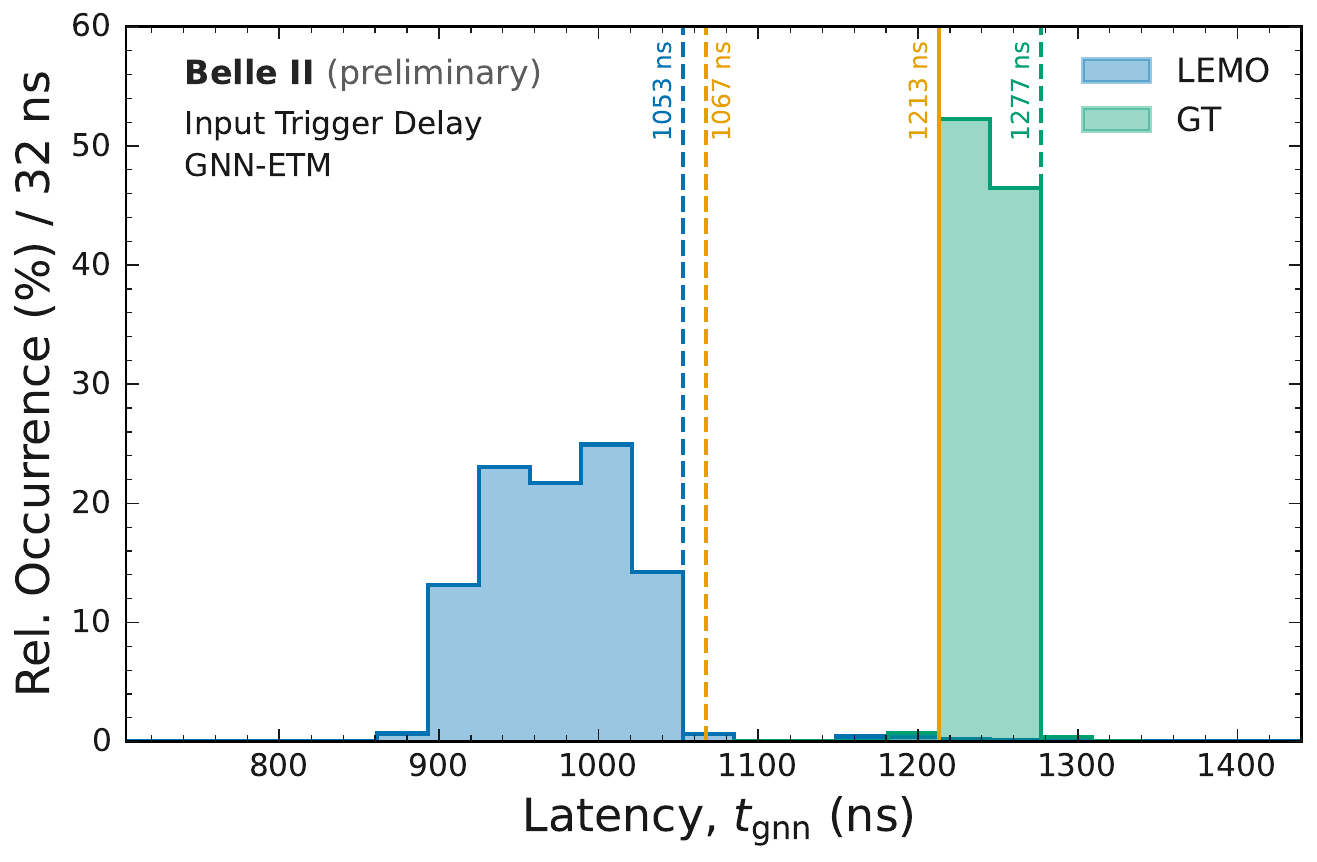}
    \caption{Relative occurrence of the \gnnetm latency $\hat{t}_\mathrm{gnn}$ of a trigger bit from \gnnetm, measured at the \gdl.
GT equals transmission via gigabit transceiver. LEMO equals transmission via twisted-pair cable.
Blue and green dashed lines denote the \SI{95}{\percent} quantile for the respective distribution.
The orange line describes the latency requirement for partaking in the active trigger decision with $t_\mathrm{FAM} = \SI{-146}{\nano\second}$.
The dashed orange line describes the latency requirement for partaking in the active trigger decision with $t_\mathrm{FAM} = \SI{0}{\nano\second}$.}
    \label{fig:timingmeasurement}
\end{figure}

\Cref{fig:timingmeasurement} shows the adjusted latency measurement with $t_\mathrm{sim} = \SI{1053}{\nano\second}$, and $\hat{t}_{0.95} = \SI{480}{\nano\second}$.
The twisted-pair cable meets the latency requirement.
For the configuration $t_\mathrm{FAM} = \SI{-146}{\nano\second}$, the latency margin on \gnnetm is \SI{160}{\nano\second} for trigger bits transmitted via twisted-pair cable.
Without this configuration delay offset, the latency margin shrinks to \SI{14}{\nano\second}.
In both configurations, the latency of the \gnnetm is too high to transmit trigger bits over gigabit transceivers, due to the overhead of the physical communication layer.

Three latency bounds can therefore be derived from experimental measurements:

\begin{enumerate}
  \item In the current configuration, a latency bound of \SI{1067}{\nano\second} is derived.
  \item When the programmable delay offset is applied at the Frontend Analog Module, the latency budget increases to \SI{1213}{\nano\second}. 
  \item Additionally swapping \gnnetm and \icnetm increases the latency budget further up to \SI{1367}{\nano\second}.
\end{enumerate}

To conclude, \gnnetm is ready to partake in active trigger decisions of the \belletwo L1 trigger system with a single, runtime-reconfigurable trigger bit via twisted-pair cable.

\subsection{Trigger Rate Monitoring}

In the following, we compare the trigger rates for the C2 trigger bit on the existing \icnetm and \gnnetm, using this representative trigger bit to demonstrate trigger rate monitoring.
The C2 trigger bit is a Boolean decision variable which is true if at least two clusters are detected in the ECL inner region in the \SI{250}{\nano\second} observation window of the ECL L1 trigger system.
In both systems, a per-cluster energy cut of \SI{100}{\mega\electronvolt} is applied.

\begin{figure}[h]
    \centering
    \subfloat[Cosmic run]{
        \includegraphics[width=0.98\linewidth]{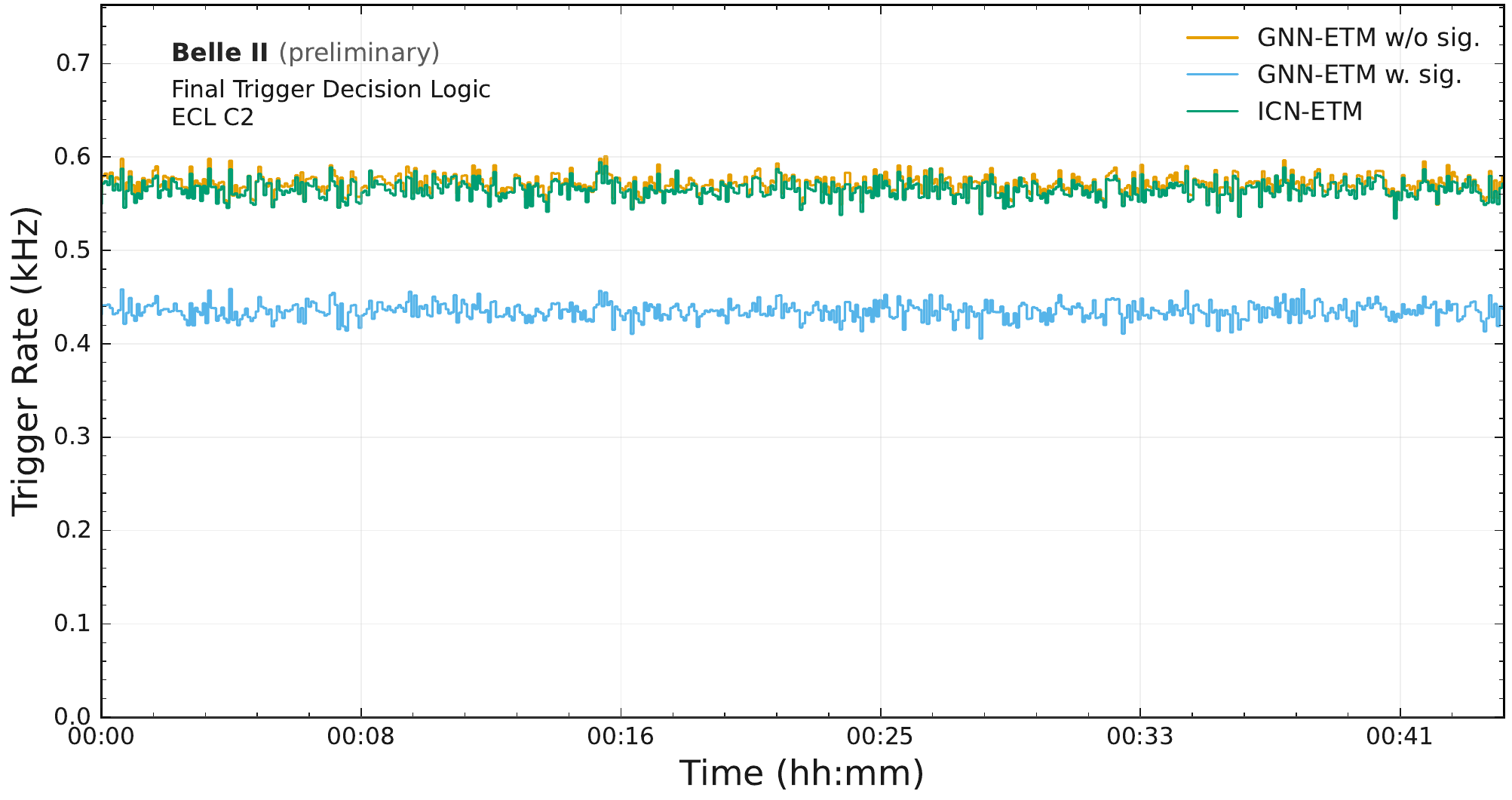}
        \label{fig:monitoring:cosmic}
    }

    \subfloat[Beam run]{
        \includegraphics[width=0.98\linewidth]{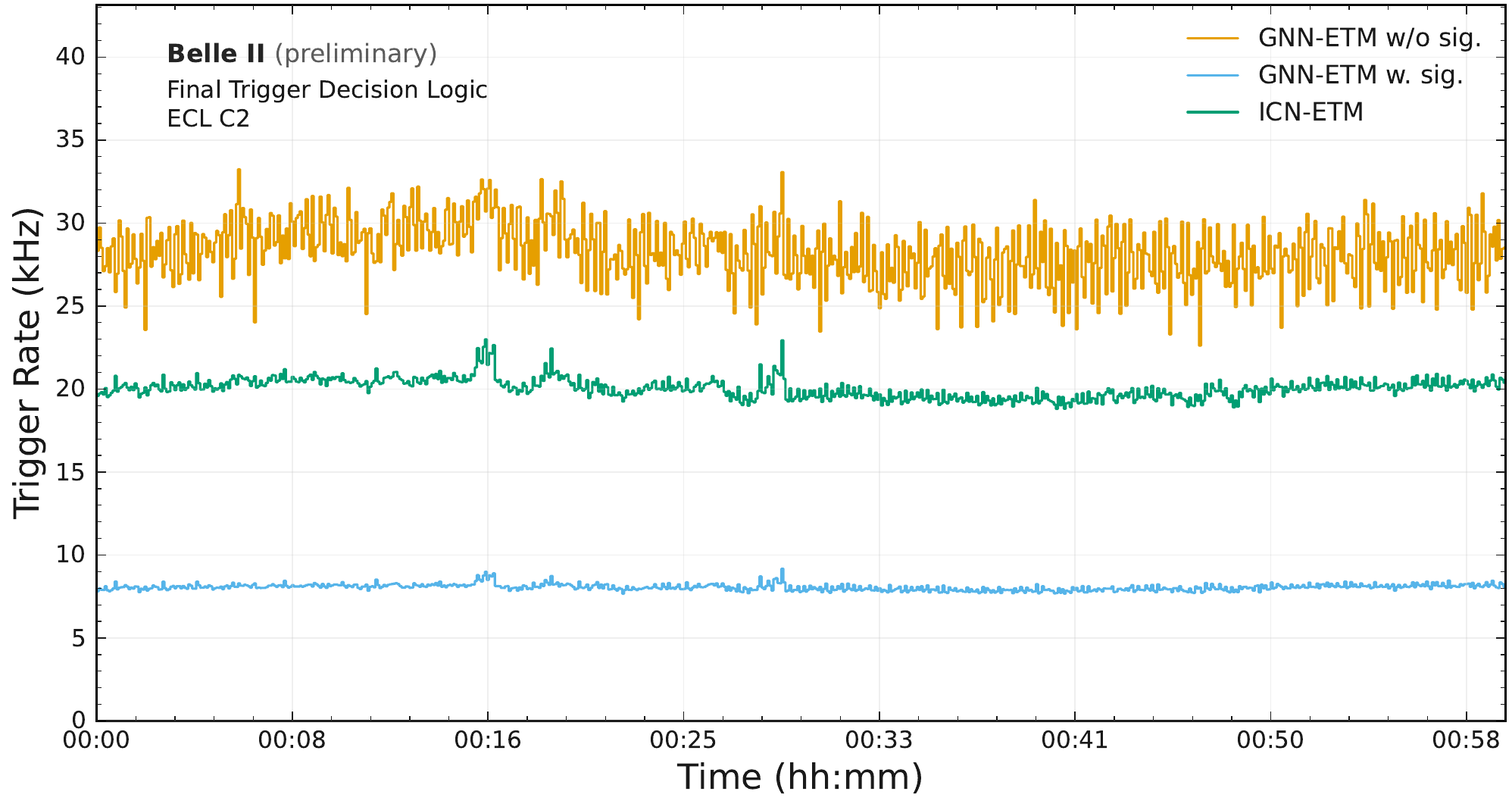}
        \label{fig:monitoring:beam}
    }
    \caption{Comparison of C2 trigger rates between \icnetm and \gnnetm based on monitoring PVs on the \gdl. \gnnetm trigger rates are shown both with the signal classifier (w. sig.) and without the signal classifier (w/o sig.).}
    \label{fig:monitoring}
\end{figure}

\Cref{fig:monitoring} shows a comparison of trigger rates between the \icnetm and the \gnnetm for the C2 trigger bit.
We select two representative runs from the \belletwo operation between May and June 2026 to demonstrate the functionality of the trigger rate monitoring.
\Cref{fig:monitoring:cosmic} shows a cosmic run without beam collisions in June 2026.
\Cref{fig:monitoring:beam} shows a physics run with beam collisions in May 2026.
The rates are based on the Final Trigger Monitor PVs on the \gdl after applying both the Bhabha and injection vetoes.
As a baseline, we depict the trigger rate of the existing \icnetm C2 trigger bit.
For the \gnnetm, we show two versions:
First, we show the C2 trigger bit based on the clusters selected by the condensation-point selection algorithm without consideration of the signal classifier per-cluster output~\cite{haide:2026}.
This trigger rate is denoted as \gnnetm w/o sig.
Second, we show the same C2 trigger bit, but we apply the signal classifier to each cluster.
As a result, background clusters are masked, and the total number of clusters per detector snapshot decreases, so the threshold of at least two clusters is reached in fewer cases.
This trigger rate is denoted as \gnnetm w. sig.

In \Cref{fig:monitoring:cosmic}, we observe that the C2 trigger rates of the \icnetm and the \gnnetm without a signal classifier are almost identical. 
Applying the signal classifier on \gnnetm reduces the C2 trigger rate by approx. \SI{100}{\hertz}.
In \Cref{fig:monitoring:beam}, we observe a higher C2 trigger rate for the \gnnetm w/o sig.\ in comparison to the \icnetm.
A potential cause of this increased rate is the ability of \gnnetm to split energy depositions into multiple clusters.
Another potential cause is the characteristic of the \icnetm, to shift the position of a cluster towards the forward endcap due to the way TCs are defined in the inhomogeneous endcap region.
Thus, the \icnetm is more likely to have only one cluster in the ECL inner region, as required by the C2 trigger bit, which leads to a lower trigger rate than the \gnnetm.
After applying the signal classifier, the trigger rate of \gnnetm drops significantly below the \icnetm rate.

In general, both online measurements of the \gnnetm and the \icnetm confirm the trends presented in ref.~\cite{haide:2026}:

\begin{enumerate}
    \item For cosmic runs without beam background, the cluster-finding performance of \icnetm and \gnnetm w/o sig. is almost identical.
    \item The \gnnetm signal classifier can significantly reduce the trigger rate.
    \item For beam runs, a greater difference in trigger rates is expected between \gnnetm and \icnetm.
\end{enumerate}

Nevertheless, a more thorough analysis is required to make quantitative statements on the two systems.
By making online monitoring available, this work lays the groundwork for a quantitative comparison of the two modules
in the \belletwo ECL trigger system.

\section{Conclusion}
\label{sec:conclusion}
In this work, we have presented the commissioning and low-latency operation of the
\gnnetm, a GNN-based calorimeter clustering trigger algorithm, in the \belletwo experiment.
For the commissioning of the \gnnetm in the L1 trigger system, we have reduced the
end-to-end latency of the FPGA-based system by a factor of \num{3}
from \SI{3168}{\nano\second} to \SI{1053}{\nano\second}.
In addition, we have integrated trigger-bit generation into the postprocessing stage and
completed the connection between the \gnnetm and the \gdl.
We confirm in a measurement that the end-to-end latency target of \SI{1067}{\nano\second}
is met with a margin of \SI{14}{\nano\second} during operation.
Structural modifications of the \belletwo ECL L1 trigger relax the latency budget to up to
\SI{1367}{\nano\second}, enabling the integration of more complex trigger bits.
In addition, the logging of trigger rates in the \belletwo EPICS database enables a
quantitative comparison of the \gnnetm and the \icnetm in a physics analysis.

\bibliographystyle{IEEEtran}
\bibliography{biblio.bib}

\end{document}